\documentclass[twocolumn]{aastex62}
\usepackage{graphics,epsf}
\usepackage{amsmath}                
\usepackage{amsfonts}               
\usepackage{amssymb}                
\usepackage{epsfig}                 
\usepackage{graphicx}
\usepackage{float}
\usepackage{color}
\usepackage[para,online,flushleft]{threeparttable}

\newcommand{\km}{{~\rm km}}
\newcommand{\s}{{~\rm s}}

\newcommand{\erg}{{~\rm erg}}

\begin{document}

\title{Evaporating planets in type Ia supernovae}



\author[0000-0003-0375-8987]{Noam Soker}
\affiliation{Department of Physics, Technion, Haifa, 3200003, Israel; soker@physics.technion.ac.il}


   


\section{Small hydrogen mass in SNe Ia  }
\label{sec:intro}

There are recent claims for the presence of hydrogen, at low abundances, in the ejecta of two type Ia supernovae (SNe Ia), which I claim is compatible with the evaporation of Jupiter-like (gas giant) planets.  

\cite{Kollmeieretal2019} used their detection of an H$\alpha$ emission line in the nebular phase to estimate the hydrogen mass in the sub-luminous SN Ia SN2018fhw/ASASSN-18tb to be $M_{\rm H} \approx 2 \times 10^{-3} M_\odot$, and possibly up to $ M_{\rm H} \approx 0.01 M_\odot$.  
For the origin of the H$\alpha$ line they considered stripped gas from a non-degenerate star in the single-degenerate (SD) scenario of SNe Ia, interaction with circumstellar matter (CSM; also \citealt{Vallelyetal2019}), or fluorescent UV pumping in a slowly expanding shell of material. 
\cite{Vallelyetal2019} commented that this SN is very different from other known CSM-interacting SNe Ia. 

\cite{Prietoetal2019} reported the detection of a narrow H$\alpha$ emission line, FWHM$\approx 1200 \km \s^{-1}$, in the low-luminosity fast declining SN Ia SN2018cqj/ATLAS18qtd. 
They inferred a hydrogen mass of $M_{\rm H} \approx 10^{-3} M_\odot$ and 
argued that the H$\alpha$ line properties are consistent with stripped hydrogen. However, the inferred hydrogen mass is significantly less than what theoretical calculations (e.g., \citealt{Botynszkietal2018}) give for the classical SD scenario. Instead, I will consider a planetary origin for this hydrogen. 
 
\section{Ejecta-evaporated plants}
\label{sec:Evaporating}

I consider the possibility that in rare cases the SN Ia ejecta unbinds a gas giant planet of mass $M_{\rm p} \simeq 1-15 M_{\rm J}$ and thus enriches the ejecta with a hydrogen mass of $M_{\rm H} \approx 1-10 M_{\rm J}$, where $M_{\rm J}$ is Jupiter mass. 

The binding energy of a Jupiter-like planet is (e.g., \citealt{LecavelierDesEtangs2007}, and adding internal energy) $E_{\rm bin} \approx 2 \times 10^{43} (M_{\rm p}/M_{\rm J})^2 (R_{\rm p}/R_{\rm J})^{-1} \erg$. In the relevant mass range the planet radius is $R_p=R_J$, where $R_J$ is the radius of Jupiter. 

The SN ejecta hits the planet and transfers a fraction $\eta \simeq 0.02$ \citep{Boehneretal2017} of its kinetic energy, $E_{\rm exp} \simeq 10^{51} \erg$,  to evaporate the planet. For the ejecta to destroy the planet, the plant must be within a distance of of 
\begin{equation}
a_{\rm max}  \simeq  50   
\left( \frac{\eta}{0.02} \right)^{1/2} 
\left( \frac{M_{\rm p}}{M_{\rm J}} \right)^{-1}  
\left( \frac{R_{\rm p}}{R_{\rm J}} \right) ^{3/2}
R_\odot.
\label{eq:Amax}
\end{equation}
 
\section{Accounting for planets at explosion}
\label{sec:Scenarios}
  
Observations and their interpretations suggest that many white dwarfs (WDs) host planetary systems (e.g., \citealt{Veras2016}), including post-common envelope evolution (post-CEE) WDs in close binary systems (e.g., \citealt{ZorotovicSchreiber2013}). However, these post-CEE circumbinary planets have a separation too large to explain complete planet evaporation by the SN ejecta. 
  
The evaporated planets might be first generation planets, i.e., planets born with the main sequence progenitor of the WD (e.g., \citealt{BearSoker2014}), or second generation planets, i.e., planets born from the mass lost by the giant progenitor of the WD (e.g., \citealt{Perets2010}). Of particular relevance to SN Ia scenarios that involve a CEE is the possible formation of a post-CEE circumbinary disk \citep{KashiSoker2011} where planets might form (e.g., \citealt{SchleicherDreizler2014}).  

Table 1 presents my estimate of the likelihood of formation of a planet at a separation of $a \la 50 R_\odot$ in each of the five SN Ia binary scenarios (for a summary of the five SN Ia scenarios and those that experience the CEE see \citealt{Soker2019}).  
\begin{table*}
\scriptsize
\begin{center}
  \caption{Origin and masses of rare planets at $a<a_{\rm max}$ in SNe Ia}
    \begin{tabular}{| p{2.2cm} | p{2.5cm}| p{2.5cm}| p{2.5cm}| p{2.5cm} | p{2.5cm} |}
\hline  
{Scenario}  & {Core Degenerate}    & {Double Degenerate} & {Double Detonation} & {Single Degenerate} & {WD-WD collision} \\
\hline  
{Relevant scenario's ingredients (CEE stands for common envelope evolution)}  
& A CO/HeCO WD experiences a CEE and merges with the CO/HeCO core of the giant. 
& A CO/HeCO WD survives a CEE close to the CO/HeCO core remnant.
& A WD or a He star survive a CEE close to the core remnant. One of the stars has a He layer.
& (1) A main sequence star experiences a CEE and survives. (2) A mass-donor giant. No CEE. 
& Two wide CO WDs evolve independently and merge long after their formation.  \\
\hline  
{$1^{\rm st}$ generation orbiting only the exploding WD(s); $M_{\rm p} \approx 5-15 M_{\rm J}$.}
  & No.
  & No.
  & No.
  & Only when the mass-donor is a giant, but unlikely.
  & Likely in 1 or 2 of the colliding WDs. \\
\hline  
{$1^{\rm st}$ generation circumbinary; $M_{\rm p} \approx 5-15 M_{\rm J}$.}
  & Likely. 
  & Likely. 
  & Likely. 
  & Likely in the case of a main sequence mass-donor.
  & No.  \\
\hline  
{$2^{\rm nd}$ generation orbiting only the exploding WD(s); $M_{\rm p} \approx 1-3 M_{\rm J}$.}
  & No.
  & No.
  & No.
  & Only when the mass-donor is a giant, but unlikely.
  & Possible in 1 or 2 of the colliding WDs. \\
\hline  
{$2^{\rm nd}$ generation from a circumbinary post-CEE disk; $M_{\rm p} \approx 1-3 M_{\rm J}$. 
}
  & Possible. 
  & Possible.  
  & Possible.  
  & Possible in the case of a main sequence mass-donor.
  & No. \\
\hline  
{Comments}
  & The planet orbits a single WD at explosion. 
  & A massive planet might influence the evolution of the two WDs to merge. 
  & A massive planet might influence the evolution of the He-rich star toward the WD. 
  & Low $M_{\rm H}$ implies a long delay from accretion to explosion (the spin-up/spin-down channel).
  & The evolution to form a low mass single WD favours the survival of a planet in a CEE.  \\
\hline  
 \hline  

     \end{tabular}
  \label{tab:Table1}\\
\end{center}
\begin{flushleft}
Notes:\\ {The term `Likely' in the second and third rows and the term `Possible' in the fourth and fifth rows imply that if a planet does exist in one of these SNe Ia scenarios, it is more likely to be a first rather than a second generation planet.  CO/HeCO means that the WD can be a CO WD or a HeCO WD.} 
\end{flushleft}
\end{table*}

For a first generation planet to be in a separation of $a \la a_{\rm max}$ at explosion, in most scenarios it should survive a CEE phase. For that, its mass must be $M_{\rm p} \ga 5 M_{\rm J}$ \citep{LivioSoker1984}, and the mass of the giant envelope cannot be too large, otherwise the planet inspirals to the core and the core destroys the planet. Note that in scenarios that suffer a CEE the planet enters the CEE together with the binary system. We actually have three bodies inside an envelope: (1) A WD; (2) Another WD or a  main sequence star; (3) a planet. In the case of the WD-WD collision scenario only the planet enters the CEE as the two WDs are in a wide binary (or triple or quadruple) system.  
 
It is not clear whether a second generation planet at a separation of $a \la a_{\rm max}$ can form. This is a close distance from a very bright WD remnant of an asymptotic giant branch star, or from a core-WD merger product in the case of the core-degenerate scenario. However, the many gas giant planets at close orbits around main sequence stars suggest that this is possible, at least in principle. 
Like the entire speculative suggestion of this short article, the formation of second generation planets at a separation of $a < a_{\rm max} \approx 50 R_\odot$ should also be a subject of future study.

In any case, the presence of a planet, whether a first or a second generation planet, requires a not too violent CEE. This requirement might prefer sub-luminous SNe Ia.

\end{document}